\documentclass[final,journal,letterpaper]{IEEEtran}

 \usepackage[hidelinks]{hyperref}
 \usepackage{varioref} 
 \usepackage{wrapfig} 
 \usepackage{threeparttable} 
 \usepackage{dcolumn} 
  \newcolumntype{d}{D{.}{.}{-1}}
 \usepackage{nomencl} 
  \makeglossary
 \usepackage{subfigmat} 
 \usepackage{fancyvrb} 
  \fvset{fontsize=\footnotesize,xleftmargin=2em}
 \usepackage{lettrine} 

 \usepackage{epsfig}
 \usepackage{epstopdf}
 \usepackage{subfigure}
 \usepackage{latexsym}
 \usepackage{amsmath}
 \usepackage{amsfonts}
 \usepackage{amssymb}
 \usepackage{mathrsfs}
 \usepackage{cases}
 \usepackage{array}
 \usepackage{color}
 \usepackage{soul}
 \setulcolor{red}
 \sethlcolor{yellow}

 \usepackage{url}
 \usepackage[noadjust]{cite}
 \usepackage{graphicx}
 \usepackage{psfrag}
 \usepackage{color}
   \usepackage{multirow}
 \usepackage{ragged2e}
 \usepackage{graphicx}  
 
 \usepackage{algorithm,algorithmic}
 \usepackage{breqn}
 \newcommand{\comment}[1]{}
 \renewcommand\color[1]{#1}

\begin{document}
%
\title{Accelerated Deep Reinforcement Learning Based Load Shedding for Emergency Voltage Control}

%
%
%

\author{Renke Huang,~\IEEEmembership{Senior Member,~IEEE,}
        Yujiao Chen, Tianzhixi Yin, Xinya Li, \\Ang Li, Jie Tan,   Wenhao Yu, Yuan Liu,~\IEEEmembership{Member,~IEEE}, Qiuhua Huang,~\IEEEmembership{Member,~IEEE}

\thanks{Pacific Northwest National Laboratory (PNNL) is operated by Battelle for the U.S. Department of Energy (DOE) under Contract DE-AC05-76RL01830. This work was funded by DOE ARPA-E OPEN 2018 Program. (\textit{Corresponding author: Qiuhua Huang})}
\thanks{R. Huang, T. Yin, X. Li, Y. Liu, A. Li and Q. Huang are with PNNL, Richland, WA 99354, USA (e-mail: \{renke.huang,  tianzhixi.yin, xinya.li,  ang.li, yuan.liu, qiuhua.huang\}@pnnl.gov).}
\thanks{Y. Chen was a post-doc with PNNL  (e-mail: yujiao.chen@pnnl.gov ).} 
\thanks{J. Tan and W. Yu are with Google Brain, Google Inc, Mountain View, CA, 94043 USA (e-mail: \{jietan, magicmelon\}@google.com).}}

%
%

%

\maketitle

\begin{abstract}
Load shedding has been one of the most widely used and effective emergency control approaches against voltage instability. With increased uncertainties and rapidly changing operational conditions in power systems, existing methods have outstanding issues in terms of either speed, adaptiveness, or scalability.  Deep reinforcement learning (DRL) was regarded and adopted as a promising approach for fast and adaptive grid stability control in recent years. However, existing DRL algorithms show two outstanding issues when being applied to power system control problems: 1) computational inefficiency that requires extensive training and tuning time; and 2) poor scalability making it difficult to scale to high dimensional control problems. To overcome these issues, an accelerated DRL algorithm named PARS was developed and tailored for power system voltage stability control via load shedding. PARS features high scalability and is easy to tune with only five main hyperparameters. The method was tested on both the IEEE 39-bus and IEEE 300-bus systems, and the latter is by far the largest scale for such a study. Test results show that, compared to other methods including model-predictive control (MPC) and  proximal  policy  optimization(PPO) methods, PARS shows better computational efficiency (faster convergence), more robustness in learning, excellent scalability and generalization capability.

\end{abstract}


\section{Introduction}

%
%
%
%


\subsection{Motivation}\label{sec:intro}

\IEEEPARstart{B}{ulk} power systems are facing increasing risks of voltage instability with greater presence of dynamic loads and expanding integration of inverter-based resources that lead to lower system strength and limited reactive capability during disturbances. Furthermore, tripping of distributed and centralized inverter-based resources during large transmission disturbances could deteriorate voltage stability and lead to voltage collapse or blackout \cite{AEMO2017report}. Load shedding has been one of the most widely used and effective emergency control approaches to counteracting voltage instability, in particular short-term voltage instability \cite{van2007voltage,van2007vs_overview,huang2019loadshedding_DRL}. {\color{blue} On the one hand, short-term voltage stability events are fast and usually last 5 to 30 seconds. Thus, it requires fast solutions and fast actions (e.g., less than half a second for each control interval\cite{Dong2017_EDR_LS}) to control and mitigate them in real-time operation. On the other hand, effectively determining load shedding time, location and amount for large power system is a complex, dynamic, and sequential decision-making problem\cite{huang2019loadshedding_DRL}}. However, existing methods have outstanding issues in terms of either speed, adaptiveness, or scalability. Thus, major enhancements of voltage control schemes are much needed. This paper focuses on developing an accelerated deep reinforcement learning (DRL)-based control method to make load shedding for emergency voltage control fast, adaptive, and scalable. 

To date, it takes extensive time (ranging from tens of hours to days ) to train a practically good control policy for even {\color{blue} small-scale} power systems using existing state-of-the-art DRL algorithms \cite{huang2019loadshedding_DRL}. Accelerated training not only makes DRL more practical for real-world large-scale power grid control applications, but also brings significant benefits, including: 1) shorter experiment turnaround time facilitating better control design  \cite{stooke2018accelerated}; 2) overcoming the operational challenges associated with increasing uncertainties by enabling DRL training closer to real-time situations (e.g., moving from days ahead to hours ahead); and 3) the ability to update the emergency control schemes more frequently whenever necessary, which can enhance the adaptiveness and effectiveness of emergency control schemes during fast-changing events such as hurricanes and cascading outages. 

\subsection{Related Work}\label{sec:related_work}
In designing a load shedding control scheme, the time, location, and amount are important and
closely related aspects of load shedding against voltage instability \cite{van2007voltage}. Past efforts in determining these aspects for load shedding-based emergency control can be roughly categorized into rule-based, measurement-based, model-based, and learning-based (including data-driven) approaches.

\textit{Rule-based approaches}: Most existing load shedding control designs deployed in the power industry are rule-based, {\color{blue} which could provide the load shedding control actions in a very short time since the rules are determined based on off-line studies and do not change online for different operation conditions or faults.} A simple scheme relies on simple rules like ``if
voltage drops below some threshold $V_{th}$ for some duration $\tau$,
shed some power $\Delta P$'' \cite{van2007vs_overview}. In addition, the settings of the rules tend to be conservative. While some enhancements to simple rule-based methods were proposed \cite{Lefebvre2003ls4HQ}, the main issues of such methods are lack of adaptiveness and optimality. {\color{blue} The main reason is that rule-based methods do not consider the spatial and temporal sensitivity and correlation of load shedding actions with respect to the effectiveness of recovering the system voltages, which is critical for achieving effective and adaptive emergency controls.  This spatial and temporal sensitivity and correlation of load shedding actions vary over time for different operation conditions and contingencies, thus it cannot be captured by fixed rules. } 

\textit{Model-based approaches}: Security-constrained alternating current optimal power flow (AC-OPF)  \cite{Misra2017} was proposed for grid emergency control. Another widely used model-based approach is model-predictive control (MPC){\color{blue} \cite{jin2009model,amraee2011adaptive}. MPC could provide optimal or near-optimal solutions for dynamic, sequential load shedding optimization problems. However, model-based methods suffer from poor scalability and long solution time issues for large-scale grid control problems, due to the computational complexity. Consequently, they can not meet the short response time requirement of grid emergency control (usually less than half a second \cite{Dong2017_EDR_LS}).} In addition, since they rely on a system model, these methods are susceptible to model inaccuracies\cite{huang2019loadshedding_DRL}.

\textit{Measurement-based approaches}: In recent years, methods for real-time voltage control were developed by leveraging  phasor measurement unit technologies and methodologies have been developed for tracking voltage behavior \cite{Glavic2012VC,Amar2019pmu,Sun2019review_voltage_control}. {\color{blue}The measurement-based approach can estimate the “sensitivity” of local load shedding actions to determine the appropriate load shedding amount for each local load. In \cite{Glavic2012VC} an online load shedding strategy by estimating the motor kinetic energy using PMU voltage and current measurements is proposed to determine the load shedding amount. In \cite{Amar2019pmu}, the load shedding actions that ensure voltage recovery within a pre-specified time are determined by an index calculated based on active and reactive power measurements.} The measurement-based method could meet the response time requirement for emergency controls, since the proposed algorithms for estimating the sensitivity indices typically have a lower computation complexity when compared with model-based methods. However, without high-level coordination that would require some model-based or learning-based approaches, these methods may not be adequate to mitigate emergencies on large-size power grids.

\textit{Learning-based approaches}: Learning-based (or data-driven) methods gained much attention and interest for grid control in both academia and industry in the past decade. {\color{blue}The basic idea of learning-based methods is to learn a (usually non-linear) mapping between the system operating conditions and  control actions from data (mostly collected from simulation), and the mapping is encapsulated in machine learning models such as a neural network. During the online application, the trained  model can infer control decisions within a very short time based on the actual system conditions.} A decision-tree-based approach was proposed for preventive and corrective control \cite{Genc2010_DT}. A hierarchical, extreme learning machine-based method for load shedding against fault-induced delayed voltage recovery (FIDVR) events was developed in \cite{Qiao2020hierarchical_LS}. An important learning-based approach for controlling dynamic systems is reinforcement learning (RL). There is a significant number of previous efforts in utilizing conventional RL methods such as Q-learning for many different power system control applications and areas \cite{Glavic2017Review,glavic2019DRL4GC}. Yet conventional RL approaches have serious limitations in terms of processing high-dimensional observation and action spaces as well as difficulty in large-scale training. These limitations are largely overcame by recent advancement of DRL algorithms, which have led to many breakthroughs in controlling complex systems, particularly in  games, robotic control, and  autonomous driving. In our previous work \cite{huang2019loadshedding_DRL}, we adapted a popular DRL algorithm called deep Q-network (DQN) and achieved adaptive emergency control schemes for both generator dynamic breaking and under-voltage load shedding (UVLS). Another DRL algorithm, deep deterministic policy gradient (DDPG), was applied for emergency load shedding schemes in \cite{zhang2018loadshedding_DRL}. DDPG has also been applied for power grid emergency frequency control in \cite{yan2020multi, chen2020model}.

It should be noted that many state-of-the-art DRL algorithms such as DQN and DDPG used in \cite{zhang2018loadshedding_DRL,huang2019loadshedding_DRL, yan2020multi, chen2020model} are notoriously known for difficulty in scaling up the solutions, and time-consuming for training and hyper-parameter tuning. However, the training (including tuning) time and scalability issues were not addressed in the previous power grid domain applications \cite{zhang2018loadshedding_DRL,huang2019loadshedding_DRL,yan2020multi, chen2020model}. {\color{blue}As a result, application of DRL-based power grid emergency control was limited to small-size power grids such as the IEEE 39-bus or 68-bus test systems.} 

Techniques for accelerating some state-of-the-art policy-gradient and value-based DRL algorithms, particularly by leveraging a combination of central and graphical processing units (CPUs and GPUs), were developed in \cite{stooke2018accelerated, liang2018rllib}. It should also be noted that these techniques and their implementations are mainly targeted and optimized for video game environments, which can be efficiently simulated and processed in GPUs. 
{\color{blue}There are two notable differences in power system control applications:  Firstly, power system dynamic simulation is computationally intensive, and the tools are CPU-based. 
Secondly, levels of difficulty in emergency control for different operating conditions and contingency events could be significantly different. 
The former means that the training should be done on CPUs and efficiently performing the power system dynamic simulation in parallel and integrating it with the DRL algorithm is the key for speeding up the training; while the latter indicates that it is also critical to have proper mechanisms to make good balance of difficulty levels of the tasks sampled for each iteration in training.} 
Consequently, existing accelerated techniques and frameworks \cite{stooke2018accelerated, liang2018rllib} are generally not well suited for power system control applications.

\subsection{Contributions}
{\color{blue} The main contributions of our previous work\cite{huang2019loadshedding_DRL} are 1) development of the first open-sourced tool for developing and benchmarking reinforcement learning algorithms for grid dynamic control, and 2) adaption of the DQN algorithm for two grid emergency control schemes. Through our previous work and the literature, we noticed some notable issues of prolonged training and tuning time and poor scalability with the DQN algorithm and many other existing DRL algorithms. Both issues pose grand challenges for developing and applying DRL-based control applications for large-scale power systems. In this paper, we aim to tackle both issues through developing a novel, much more scalable, yet easy to tune DRL algorithm.} The main contributions of this paper include:
\begin{itemize}
    \item {\color{blue}Novel and significant improvements of the ARS algorithm \cite{mania2018simple} with deep neural network architectures like the forward neural network (FNN) and the long short-term memory (LSTM) network. The improvements keep the advantages of the ARS algorithm in supporting large learning rates and having only a few hyper-parameters to tune, meanwhile significantly enhancing its capability of tackling large-scale, high-dimensional, and highly nonlinear grid control problems with the support from FNN and LSTM.}
    \item Accelerating the ARS algorithm for grid control with a novel scheme for integrating and parallelizing the ARS algorithm and power system dynamic simulations systematically, leading to 136X speedup training on the IEEE 300-bus system.
    \item {\color{blue}To our best knowledge, our work is the first to apply this Parallel ARS algorithm enhanced with FNN and LSTM for power grid emergency control. We have compared our proposed method with both the model-based MPC method and the parallel proximal policy optimization (PPO) method which is one  state-of-the-art DRL method implemented in RLlib \cite{liang2018rllib}. As shown in Section \ref{subsec:IEEE300}, our proposed method outperformed MPC in solution time by two orders while yielding comparable results. When compared with the parallel PPO method, our method  achieved faster convergence and better solutions while using much less training time.}
    \item {\color{blue} Our results reveal that with novel enhancements and proper adaptation, conventionally conceived worse performance of derivative-free methods such as the ARS method can become an outstanding approach for solving complex grid control problems, even when compared to commonly believed more sample efficient methods such as PPO. The results highlight the importance of  considering multiple aspects of DRL algorithms and requirements of grid control applications when selecting DRL methods and the necessity of comprehensively benchmarking them for various grid control problems in an open and transparent manner.} 
\end{itemize}

\subsection{Organization}
The rest of the paper is organized as follows: Section \ref{sec:prob_formul} introduces DRL and the problem formulation. Section \ref{sec:ARS} presents the ARS algorithm and its enhancements and techniques for accelerating the algorithm for grid control. Test cases and results are shown in Section \ref{sec:results}.
Conclusions and future work are provided in Section \ref{sec:conclusions}.

\section{DRL-based Load Shedding for Emergency Voltage Control}\label{sec:prob_formul}

This section presents load shedding for emergency voltage control, an introduction to DRL, and the problem formulation.

\subsection{Emergency Voltage Control via Load Shedding}
\comment{

With increasing  installation  of  dynamic  loads, lower  system  strength,  and  limited  reactive  capability, transmission operators and utilities are facing higher risk of short-term voltage stability, in particular the FIDVR problem for power systems with high penetration of air-conditioning loads. FIDVR is the phenomenon that power systems experience a significantly delayed voltage recovery following a severe fault. The driving force of the delay and further system degradation (e.g.,  voltage collapse) is the unfavorable response  of  dynamic  loads  such  as  induction  motors. 
}
Among the measures of emergency voltage control, load shedding is well known as an effective countermeasure against voltage instability \cite{taylor_concepts_1992}. It has been widely adopted in the industry, mostly in the form of rule-based UVLS. The UVLS relays are usually employed to shed load demands at substations in a step-wise manner if the monitored bus voltages fall below the predefined voltage thresholds. ULVS relays have a fast response, but do not have communication or coordination between other substations, leading
to unnecessary load shedding  {\cite{Bai2011}} at affected substations.

As pointed out in the introduction, there are three key factors for load shedding: time, location, and amount. To optimally determine these three factors simultaneously, one has to solve a highly non-linear, non-convex, optimal decision-making problem. A detailed mathematical formulation as a constrained optimal control problem and its conversion to a Markov decision process (MDP) formulation can be found in the recent work of the authors \cite{huang2019loadshedding_DRL}. {\color{blue}Note that the constrained optimal control problem for load shedding control formulated in \cite{huang2019loadshedding_DRL} includes the equality constraints of dynamic differential equations of the grid, as well as inequality constraints of the dynamic states of the grid. Thus, the size of the optimization problem becomes very large for large power grids and it is computationally intensive to solve this problem with model-based MPC methods \cite{huang2019loadshedding_DRL}.} As this paper focuses on an accelerated DRL-based solution to this problem, a brief introduction to DRL is provided below, followed by a problem formulation using MDP.

\subsection{Deep Reinforcement Learning}

A RL problem can be defined as policy search in a (partially observable) MDP defined by a tuple ($\mathcal{S}, \mathcal{A}, \mathcal{P}, r$) \cite{Sutton1998}. The state space $\mathcal{S}$ and action space $\mathcal{A}$ could be continuous or discrete. In this paper, both of them are continuous. The environment transition function  $\mathcal{P} :\mathcal{S} \times \mathcal{A} \times \mathcal{S} \longrightarrow \mathbb{R}$ is the probability density of the next state  $ s_{t+1} \in \mathcal{S}$ given the current state $ s_{t} \in \mathcal{S}$ and action $ a_{t} \in \mathcal{A}$. At each interaction step, the environment returns a reward $r :\mathcal{S} \times \mathcal{A} \longrightarrow \mathbb{R}$. The standard RL objective is the expected sum of discounted rewards. The goal of an agent is to learn a policy $\pi_\theta(s_t,a_t)$ that maximizes the objective. 

DRL is a combination of RL and deep-learning technologies. The capabilities of high dimensional feature extraction and non-linear approximation from deep learning makes it possible for DRL to directly use the raw state-space representations and train policies for complex systems and tasks in a more effective and efficient way. 

\begin{figure}
\centering
\includegraphics[clip, width=0.35\textwidth]{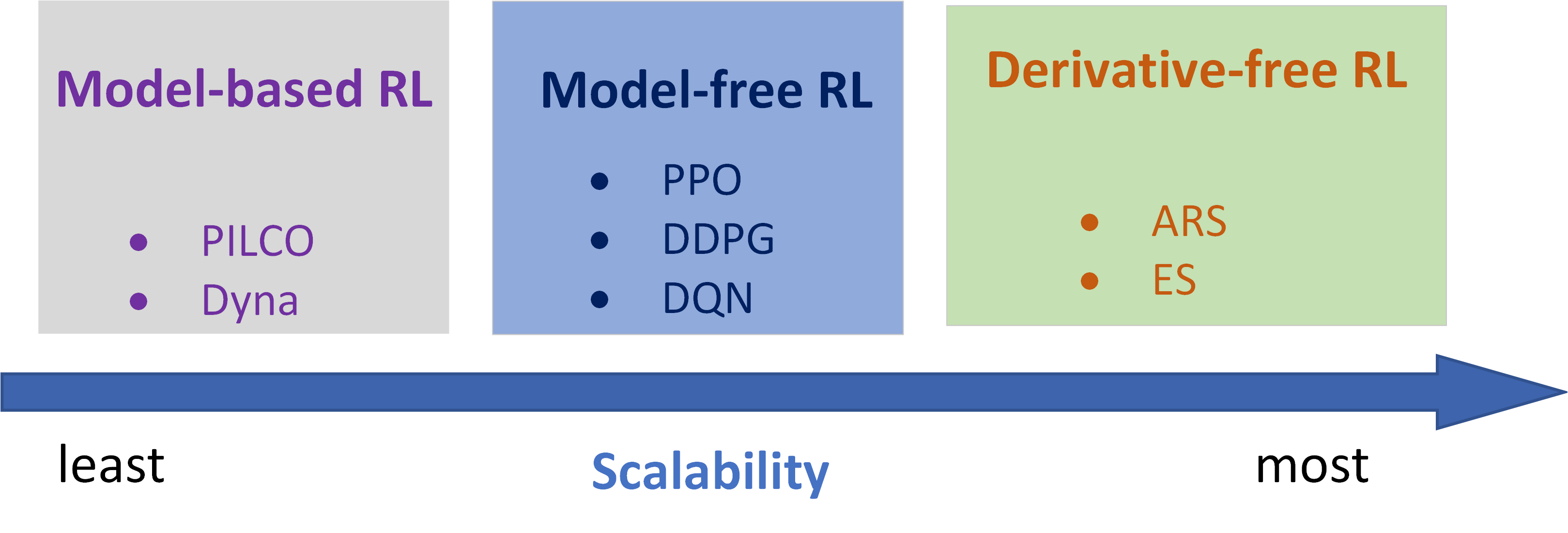}
\vspace{-3mm}
\caption{Comparison of different classes of RL algorithms}
\label{fig:RL_algo_comparison}
\vspace{-3mm} 
\end{figure}

{\color{blue}
There are three classes of RL algorithms, i.e., model-based RL, model-free RL and derivative-free RL \cite{franccois2018intro_DRL}. Fig. \ref{fig:RL_algo_comparison} shows comparisons of them in terms of computational scalability.  Computational scalability is mainly related to parallelization and speeding up the training process. In addition to scalability, main factors for determining a proper RL algorithm for one particular application include sampling efficiency and hyperparameter tuning. Sampling efficiency measures how much data need to be collected to train the RL algorithm. Regarding hyperparamter tuning, the fewer sensitive hyperparameters one RL algorithm has, the easier (and less time) it takes to tune them to achieve good performance. Generally, it is very rare that one RL algorithm has outstanding performance in all these aspects, thus the key  is to select one that can meet the requirements of grid emergency control and has good potential to overcome the bottlenecks in training time and scalability.   

For model-based RL, the planning stage has not been able to scale reliably to high dimensions, leading to the poor scalability of model-based approaches \cite{franccois2018intro_DRL}. For model-free RL, when neural networks are used as function approximation for the policy, gradients need to be calculated at some point to update the network weights during training. However, when applied for the power grid emergency voltage control problem, gradient is not easy to estimate from a sophisticated power system simulation. Recently, there are several derivative-free methods such as ARS \cite{mania2018_ARS} and natural evolution strategies (ES) \cite{salimans2017evolution} that have been developed as competitive and highly scalable alternatives to other gradient-based DRL algorithms.

Admittedly, ARS is not the most sampling (or data) efficient DRL algorithm. However, data efficiency is not the bottleneck for power grid emergency control problems because power grid simulators can be used to generate data efficiently on the fly, in parallel, and faster than real-time \cite{huang2017faster}. On the other hand, the training speed becomes a bottleneck. Many DRL algorithms need to train for days or even weeks before good policies can be learned. Such slow training time makes these methods impractical for emergency control of large-scale power grids, where both the simulation and the control policy to be learned are considerably more complex. Given the current trend that computation becomes cheaper and more computing resources are readily available, we purposefully choose a scalable learning algorithm that can take advantage of this trend. }It should be noted that the small number of sensitive hyper-parameters of ARS means it is easier to tune compared to other existing DRL methods. More details of ARS will be discussed in the next section. In light of these properties, we have adopted ARS and further enhanced and accelerated it to achieve the object of designing a fast, adaptive, and scalable DRL-based load shedding scheme in this paper, and thus overcame the bottlenecks in training speed and scalability. {\color{blue} We have successfully applied our method on the IEEE 300-bus system and reduced the training time by 136 times.
}

\comment{
The ARS algorithm generally is easier to scale in distributed settings, and thus has better computational efficiency \cite{mania2018_ARS}, while its shortcoming is the relatively poor sampling efficiency. 
As high-performance grid and parallel simulations \cite{huang2017faster} can be leveraged to fully speed up producing plentiful data to train the ARS policy offline, the weakness in sampling efficiency can be offset by its scalability advantage, provided that sufficient computing resources are available. It should be noted that the small number of sensitive hyperparameters of ARS means it is easier to tune compared to other existing DRL methods. More details of ARS will be discussed in the next section. In light of these properties, we have adopted ARS and further enhance and accelerate it to achieve the object of designing a fast, adaptive, and scalable DRL-based load shedding scheme in this paper.  
}

\subsection{MDP Formulation for Load Shedding}

\textit{State}: The observed system variables $\mathbf{O}_t$ at time $t$ include voltage magnitudes at monitored buses (denoted as $V_t$) as well as the percentage of load still remaining at controlled buses (denoted as $P_{Dt}$). To capture the dynamics of the voltage change, the most recent observed states could be stacked with some historical state records and treated as the actual state for the agent at time $t$, i.e., $s_t =(\mathbf{O}_{t-N_r-1},\cdots, \mathbf{O}_t )$.

\textit{Action}: The control action at each controlled load bus is to shed a percentage (within [0, 20\%] in this paper) of the total load at each action time step. Thus, the action space is continuous with a range of [-0.2, 0] (minus means shedding) for each action bus.

\textit{State transition}: The state transition is deterministic and governed by power system dynamics that are defined by a set of differential and algebraic equations \cite{huang2019loadshedding_DRL}. 

\begin{figure}
\centering
\includegraphics[width=0.42\textwidth]{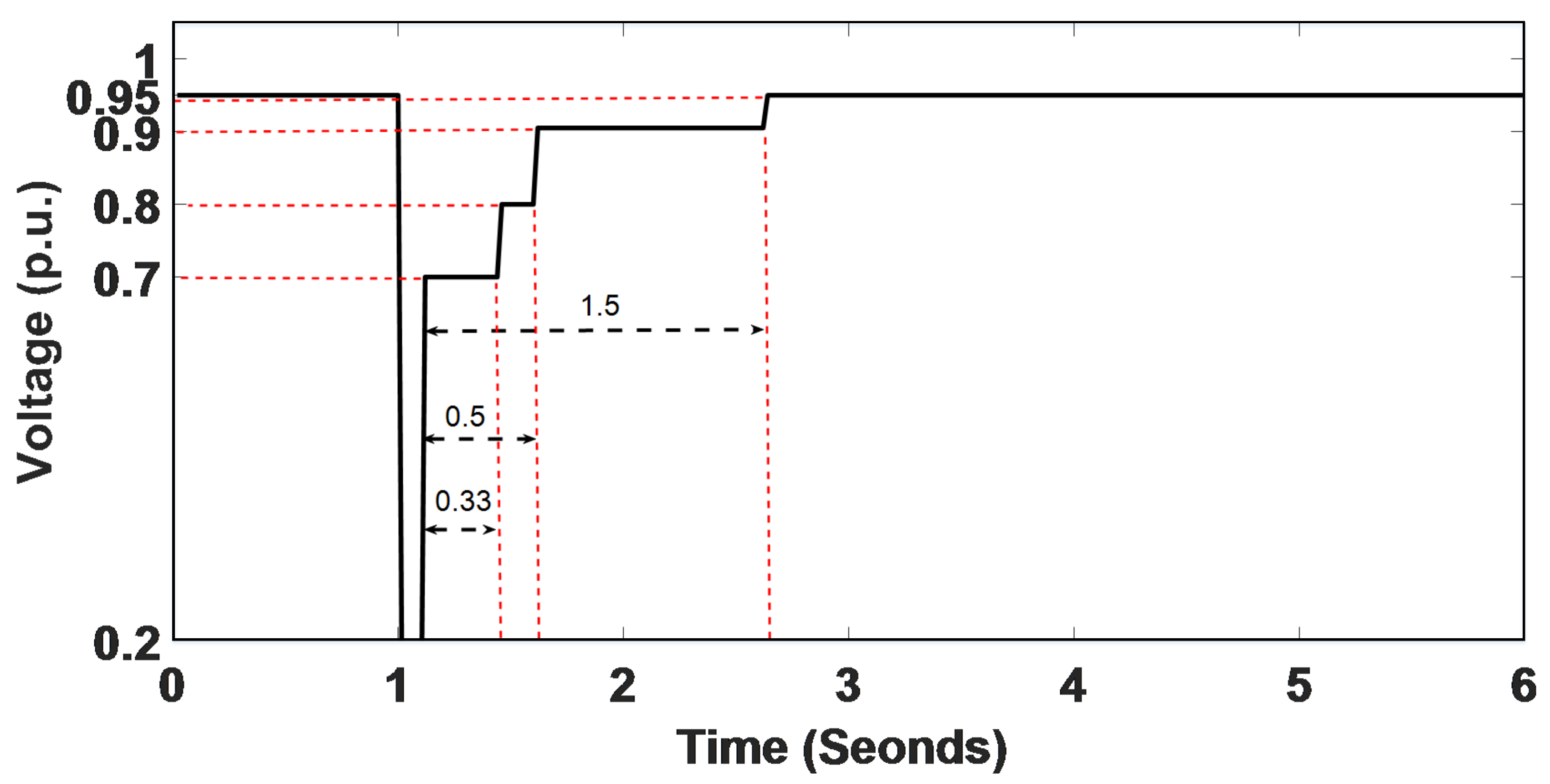}
\vspace{-3mm} 
\caption{Transient voltage recovery criterion for transmission system \cite{PJM2009}}

\label{fig:1}

\end{figure}

 \textit{Reward}: The basic principle of designing the reward function is to guide the agent to meet the transient voltage recovery criterion that is defined to evaluate the system voltage recovery. Without loss of generality, we referred to the standard proposed in \cite{PJM2009} and shown in Fig. \ref{fig:1}. After fault clearance, the standard requires that voltages should return to at least 0.8, 0.9, and 0.95 p.u. within 0.33, 0.5, and 1.5 s. Accordingly, the reward $r_t$ at time $t$ is defined as follows:
\begin{equation}\label{eq: UVLS_reward}
 r_t= 
\begin{cases}
    -M, \text{ if } V_i(t)<0.95, \quad t>T_{pf} +4		 \\
    c_1 \sum_i \Delta V_i(t) -c_2 \sum_j \Delta P_j (t) -c_3u_{ivld},             \text{ otherwise}
\end{cases}
\end{equation}
\[
\Delta V_i(t)= 
\begin{cases}
    min \left\lbrace V_i(t) - 0.7, 0 \right\rbrace , \text{ if }  {\scriptstyle T_{pf}<t<T_{pf} +0.33	}	 \\
    min \left\lbrace V_i(t) - 0.8, 0 \right\rbrace , \text{ if }  {\scriptstyle T_{pf} + 0.33<t<T_{pf} +0.5	}\\
    min \left\lbrace V_i(t) - 0.9, 0 \right\rbrace , \text{ if }  {\scriptstyle T_{pf} + 0.5<t<T_{pf} +1.5}	\\
    min \left\lbrace V_i(t) - 0.95, 0 \right\rbrace , \text{ if } {\scriptstyle  T_{pf} + 1.5<t}
\end{cases}
\]
where $T_{pf}$ is the time instant of fault clearance. The above reward function has three parts: (1) total bus voltage deviation below the standard voltage thresholds shown in Fig. \ref{fig:1}, where $V_i(t)$ is the bus voltage magnitude for bus $i$ in the power grid; (2) total load shedding amount, where $\Delta P_j (t)$ is the load shedding amount in p.u. at time step $t$ for load bus $j$; and (3) invalid action penalty $u_{ivld}$  if the DRL agent still provides load shedding action when the load at a specific bus has already been shed to zero in the previous time step, or when the system is within normal operation. The weight factors for the above three parts are $c_1, c_2,$ and $c_3$. Note that  if any bus voltage is below 0.95 p.u. 4 s after the fault is cleared, the reward is set to a large negative number {\color{blue} $-M$, which is -1,000 for the IEEE 39-bus system and -10,000 for the IEEE 300-bus system test cases in this paper.}

{\color{blue}Note that the objective of the RL algorithm is to maximize the total accumulated rewards over a specific time period $T_{max}$, i.e., $max \sum_{t=0}^{T_{max}} r_t$. We designed the reward function defined in Eqn. (\ref{eq: UVLS_reward}) in such a way that maximizing the total accumulated rewards will result in minimizing the total load shedding amount and the voltage violations at all measured buses with reference to the voltage performance envelop curve in Fig. \ref{fig:1}. In other words, the optimality of the power grid load shedding control problem is achieved by shedding as less load as possible to recover voltage buses to meet the voltage performance requirements.}

\section{ARS algorithm and its enhancements}\label{sec:ARS}
The ARS algorithm was originally proposed in \cite{mania2018_ARS} as a competitive alternative to conventional model-free DRL algorithms. In this paper, we have enhanced, accelerated, and tailored it for power system voltage stability control. 

\subsection{ARS algorithm}\label{subsec:ars_intro}

 Different from existing model-free DRL algorithms that use action-space exploration, ARS performs policy parameter-space exploration and estimates the gradient of the returns using sampled rollouts, thus back-propagation is not needed. Compared to existing DRL algorithms (i.e., TRPO, DDPG, PPO, A2C, and SAC), Mania et al. \cite{mania2018_ARS} demonstrated that ARS can achieve comparable or even better performance in robotic continuous control problems  while taking less wall clock time to train. Furthermore, in contrast to many gradient-based DRL algorithms such as DDPG and PPO having more than 20 hyperparameters,  ARS has only five main hyperparameters, i.e.,  $\alpha$, $\upsilon$, $N$, $b$, and $m$ in \textbf{Algorithm}  \ref{ARS_algo}, which makes it much easier for end users to achieve satisfactory control performance without extensive hyperparameters tuning. 
 
 The ARS algorithm employed in this paper is shown in \textbf{Algorithm} \ref{ARS_algo}, which is modified based on \cite{mania2018_ARS}. To scale up the ARS algorithm for large-scale grid control problems and reduce the training time, we proposed a novel nest parallelism scheme and implemented it on a high-performance computing platform. We also introduced a decay for step size and the perturbation noise. Details are discussed in subsection \ref{subsec:parallel_ARS}.

\setlength{\textfloatsep}{0pt}
\begin{algorithm}
 \caption{Modified ARS}
 \begin{algorithmic}[1]
  \STATE \textbf{Hyperparameters:} Step size $\alpha$, number of policy perturbation directions per iteration $N$, standard deviation of the exploration noise $\upsilon$, number of top-performing perturbed directions selected for updating weights $b$, number of rollouts per perturbation direction $m$. Decay rate $\varepsilon$.
  \STATE \textbf{Initialize:} Policy weights {$\theta_0$} with small random numbers; the running mean of observation states {$\mu_0 = \mathbf{0} \in \mathbb{R}^n$} and the running standard deviation of observation states {$\Sigma_0 = \mathbf{I}_{n} \in \mathbb{R}^n$}, where $n$ is the dimension of observation states, the total iteration number $H$.
  \FOR {iteration $t = 1,...,H$}
    \STATE sample $N$ random directions {$\delta_1, ..., \delta_N$} with the same dimension as policy weights $\theta$
    \FOR {each $\delta_i  (i \in [1,..., N])$}
    \STATE add $\pm$ perturbations to policy weights: $\theta_{ti+} = \theta_{t-1} + \upsilon \delta_i$ and $\theta_{ti-} = \theta_{t-1} - \upsilon \delta_i$
    \STATE \textbf{do} total $2m$ rollouts (episodes) denoted by $R_{p \in T}(\cdot)$ for different tasks $p$ sampled from task set $T$ based on the $\pm$ perturbed policy weights, calculate the average rewards of $m$ rollouts as the rewards for $\pm$ perturbations, i.e., $r_{ti+}$ and $r_{ti-}$  
    {
        \vspace{-3mm} 
        \begin{equation}{
        \begin{cases}
            \text{$r_{ti+} = \frac{1}{m}R_{p \in T}(\theta_{ti+} , \mu_{t-1} , \Sigma_{t-1})$} \\
            \text{$r_{ti-} = \frac{1}{m}R_{p \in T}(\theta_{ti-}, \mu_{t-1} , \Sigma_{t-1})$}\\
        \end{cases}}
        \end{equation}
        \vspace{-3mm} 
    }
    
    \STATE During each rollout, states $s_{t,k}$ at time step $k$ are first normalized and then used as the input for inference with policy $\pi_{\theta_t}$ to obtain the action $a_{t,k}$, which is applied to the environement and new states $s_{t,k+1}$ is returned, as shown in (\ref{eq: forward_pass}). The running mean $\mu_t$ and standard deviation $\Sigma_{t}$ are updated with  $s_{t,k+1}$
    {
    \vspace{-3mm} 
        \begin{equation}{
        \begin{cases}
            \text{$s_{t,k} = (s_{t,k} - \mu_{t-1})/\Sigma_{t-1}$} \\
            \text{$a_{t,k} = \pi_{\theta_t}(s_{t,k})$} \\
            \text{$s_{t,k+1} \longleftarrow \mathcal{P}(s_{t,k},a_{t,k})$} \\
        \end{cases}}
        \label{eq: forward_pass}
        \end{equation}
    \vspace{-3mm}
    }    
    \ENDFOR
    \STATE sort the directions based on $\max[r_{ti+},r_{ti-}]$ and select top $b$ directions, calculate their standard deviation $\sigma_b$
    \STATE update the policy weight:
    {
    \vspace{-3mm} 
    \begin{equation}\label{eq: weights_update}
    \theta_{t+1} = \theta_t + \frac{\alpha}{b\sigma_b}\sum\limits_{i=1}^{b}(r_{ti+}-r_{ti-})\delta_i
    \end{equation}
    \vspace{-3mm} 
    }
    \STATE Step size $\alpha$ and standard deviation of the exploration noise $\upsilon$ decay with rate $\varepsilon$: $\alpha = \varepsilon\alpha$, $\upsilon = \varepsilon\upsilon$
  \ENDFOR
 \RETURN {$\theta $}
\end{algorithmic} 
\label{ARS_algo}
\end{algorithm}

The original ARS algorithm was proposed for linear control policies with the policy $\theta$ represented by a matrix \cite{mania2018_ARS}. Our test results in Section \ref{sec:results} show that the linear policy representation does not perform well for highly non-linear power system voltage stability control problems. To overcome this shortcoming, we enhanced ARS by modeling policies with neural networks, namely the FNN and the LSTM network \cite{Schmidhuber1997_LSTM}. Details of deploying both the FNN and LSTM together with ARS are provided in subsection \ref{subsec:FNN_LSTM}. 

\subsection{Accelerating ARS Algorithms for Power System Control}\label{subsec:parallel_ARS}

To explore the parameter space of the control policy efficiently and be adaptive to multiple tasks (in this paper we define different fault scenarios in the power grid as multiple tasks, denoted by task set $T$), the ARS algorithm needs to perform a large number of power grid dynamic simulations (environment rollouts) by inferring with a sufficient number of different perturbed policies at each iteration of the training. Parallelizing power grid dynamic simulations in the ARS training plays a critical role for accelerating the training speed. ARS supports parallelism in steps 5 and 7 in \textbf{Algorithm}  \ref{ARS_algo} by nature, the crucial part is how to implement it efficiently and effectively based on the requirements and special characteristics of power system dynamic simulation and control. 

\comment{
Explain the 3-level implementation details 

1. 3-level parallelism plot and explanation. Note why  individual simulation-task level parallelism is not utilized

2. parallizing ARS + grid simulation using Ray

3. introduce the platform at a high level, highlighting integrating grid simulator into the environment and parallelizing it

4. Some key implementation details: such as sampling for rollouts, weight initialization

}

\begin{figure}[!t]
\centering
\includegraphics[width=\columnwidth ]{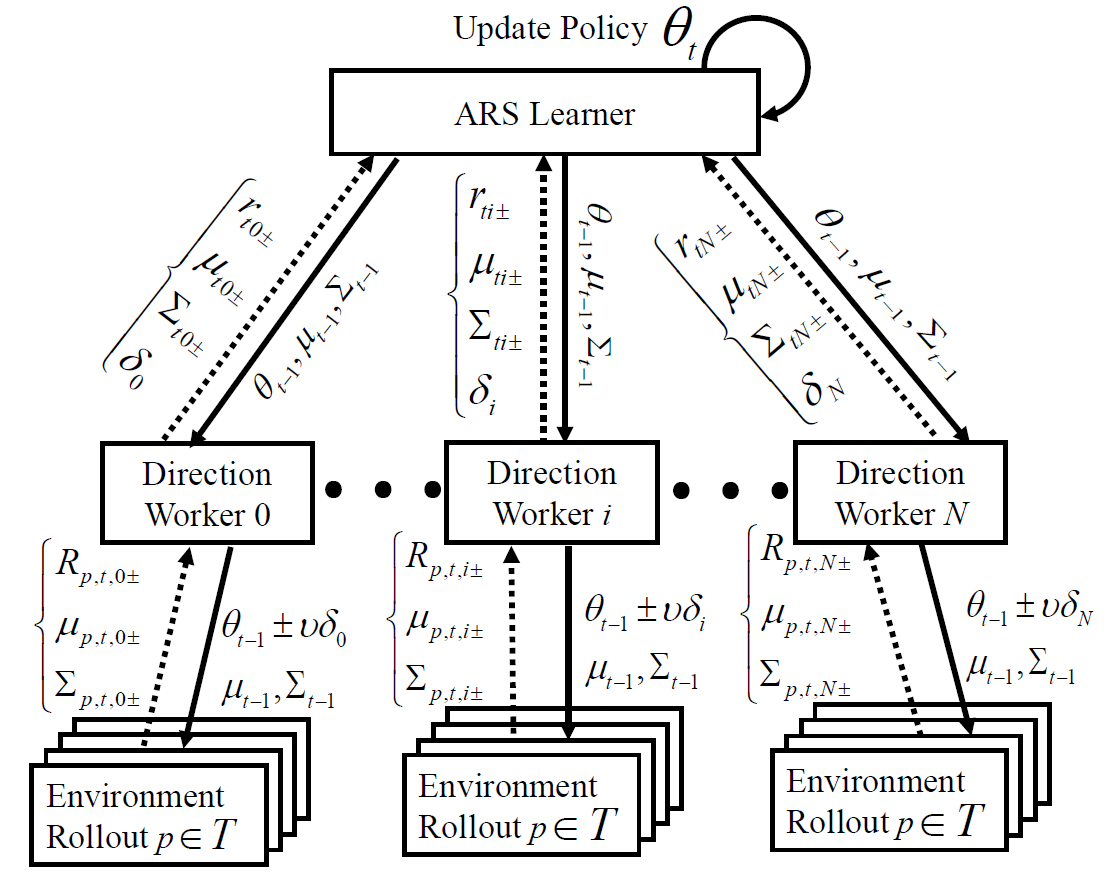}
\vspace{-3mm} 
\caption{A nest parallel scheme for accelerating the ARS algorithm}
\label{fig:np_img}
\end{figure}

The parallel version of the ARS algorithm (named PARS) is implemented with the Ray framework \cite{moritz2017ray}, which supports task parallelism (via Ray remote functions) and actor-based computation (via Ray remote classes). {\color{blue}We developed a hierarchical distributed-computing architecture, as illustrated in Fig. \ref{fig:np_img}, to fully utilize the parallelism of our problem formulation. In our formulation, both the learning algorithm and the simulation are parallelized. The first layer of our hierarchical architecture is for parallelizing the learning (step 5 in \textbf{Algorithm} \ref{ARS_algo}) and the second layer is for parallelizing the environment simulation (step 7 in \textbf{Algorithm} \ref{ARS_algo}). Although it is possible to collapse this two-layer hierarchy into one level, analogue to reshaping a 2D matrix into one vector, we deliberately choose this hierarchical structure for the following three reasons. (1) Conceptual simplicity: It is a cleaner and easy-to-maintain design to have different layers for different components (learning vs. simulation). (2) Support of different strategies for sampling tasks: it allows deliberately sampling of tasks of varied levels of difficulty in learning for each  iteration  in  training. (3) Modularity and extensibility to support full parallelism: currently in the algorithm we rollout the environment multiple times using the same perturbed policy to account for the stochasticity in the simulated environment. We also plan to further parallelize the single rollout simulation with multiple CPU cores when we apply our method on larger power grids (e.g., more than 2000 buses) in the future \cite{huang2017faster}. It is much easier to implement such extensions with the current hierarchical design: we can simply add more layers without major changes to the existing code. }

\comment {structure of the two-level parallelism is illustrated in Fig. \ref{fig:np_img}, which includes a perturbation parallelism (policy level) and an environment rollout parallelism (task level), corresponding to steps 5 and 7 in \textbf{Algorithm}  \ref{ARS_algo}, respectively. }

As shown in Fig. \ref{fig:np_img}, the ARS learner is an actor at the top to delegate tasks and collect returned information, and controls the update of policy weights $\theta$. The learner communicates with subordinate workers and each of these workers is responsible for one or more perturbations (random search) of the policy weights. The ARS learner combines the results from each worker and updates the policy weights centrally based on the perturbation results from the top performing workers. The workers do not execute environment rollout tasks by themselves. They spawn a number of slave actors and assign these tasks to subordinate slave actors. Note that each worker needs to collect the rollout results from multiple tasks inferring with the same perturbed policy, and each slave actor is only responsible for one environment rollout with the specified task and perturbed policy sent by its up-level worker. For the environment rollouts, power system dynamic simulations are performed by RLGC \cite{RLGC}, which is an open source tool for developing and benchmarking RL algorithms for grid control and supports task-level parallelism.  

Details for the data and information exchanged between the ARS learner, the workers, and the environment rollout actors are shown in Fig. \ref{fig:np_img} and described as follows. At the beginning of a training iteration $t$, the ARS learner distributes the policy weights ${\theta}_{t-1}$, the mean ${\mu}_{t-1}$, and the standard deviation ${\Sigma}_{t-1}$ of the observations from the previous iteration $t-1$ to the workers that are responsible for different policy weight perturbations. For each worker $i$, it distributes the ${\mu}_{t-1}$, ${\Sigma}_{t-1}$, as well as the specially perturbed policy weights ${{\theta }_{t-1}}\pm \upsilon {{\delta }_{i}}$ to its slave actors. Each slave actor performs single environment rollout for a different task $p\in T$ ($T$ is the set of tasks) by inferring with the perturbed policy, and sends the new mean ${{\mu }_{p,t,i\pm }}$ and standard deviation ${{\Sigma }_{p,t,i\pm }}$ of the observations as well as the reward ${{R}_{p,t,i\pm }}$ back to its master worker. Upon receiving the rewards, means, and standard deviations of observations from all slave actors, each worker $i$ computes the average reward $r_{ti\pm}$, new mean ${\mu}_{ti\pm}$, and standard deviation ${\Sigma}_{ti\pm}$ for the observations from all its tasks $T$ and sends them, together with the perturbation $\theta_{i}$, back to the ARS learner. Once the ARS learner receives all the information from its workers, it updates the policy weights according to (\ref{eq: weights_update}) in \textbf{Algorithm} \ref{ARS_algo} and the training continues to the next iteration. 

\comment{
To pass a variable, the code uses \texttt{ray.put} to store an object in the object store, and uses \texttt{ray.get} to fetch the object from the object store. The ARS learner, managers, and workers are of their own classes (ray remote actor), respectively.
}

\subsection{FNN and LSTM for Modeling Policies}\label{subsec:FNN_LSTM}

\begin{figure}
\vspace{-2mm}
\centering
\includegraphics[clip, width=0.35\textwidth]{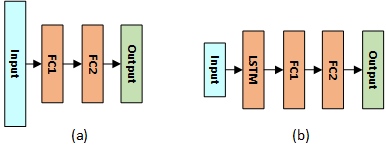}
\caption{Neural network architectures integrated with ARS: (a) FNN; (b) LSTM+FNN}
\label{fig_FNN_LSTM}
\end{figure}

We propose an innovative method of integrating the FNN and LSTM together with ARS to significantly enhance the performance of ARS. Note that traditionally weights of neural networks are updated with back-propagation using gradient descents, whereas in our PARS algorithm the weights of neural networks are updated using (\ref{eq: weights_update}) in Algorithm \ref{ARS_algo}. FNN is a commonly used neural network model for mapping the observations to actions in DRL algorithms to learn the non-linearity of their relationship. A FNN with two hidden, fully-connected (FC) layers (shown in Fig. \ref{fig_FNN_LSTM}a) is used in this paper.

\comment{

By noting $i$ the $i$\textsuperscript{th} layer of the FNN, $j$ the $j$\textsuperscript{th} hidden unit of this layer:
 \begin{equation}
 y^i = f_i((w^i_j)^T x)
 \end{equation}
\noindent where $x$ represents the input to a fully connected layer, $y$ is the output from this fully connected layer, $w$ are the learnable parameters in the network. For multiple fully connected layers, FNN is stepped from input (observation) layer to output (action) layer through the sum of weighted layer inputs after activation. 

}

\comment{
The policy weights using FNN are the network weights. For example, for a two-layer network with 64 neurons each layer (the network size is [64,64], observation size of 1540 and action size of 46, the number of policy weights to be updated is $N_w = \textbf{1540} \times 64 + 64 \times 64 + 64 \times 46 = 105600$. $R$ is an activation function used to introduce non-linear complexities to the model. In our study, hyperbolic tangent (tanh) function is used for activation:

 \begin{equation}
R(z) = \frac{e^z - e^{-z}}{e^z + e^{-z}}
 \end{equation}
 
 \noindent tanh is a recalling of the sigmoid output range from -1 to 1.
 }

 The main advantage of FNN is that its simple architecture makes it easy to train. On the other hand, lack of capability of storing historical memory makes it challenging for power system stability control applications because observed states at one step do not capture important system dynamic features such as voltage dip or recovery trend. One solution is to stack some recent history observations as the actual input to FNN. This will inevitably increase the dimension of the input, and thus size of the FNN. Furthermore, the number of history observations to be stacked becomes a hyperparameter, which needs to be tuned on a task basis. This leads us to also explore adding LSTM for modeling policies for enhancing the ARS.

\begin{figure}[!t]
\centering
\includegraphics[width=\columnwidth ]{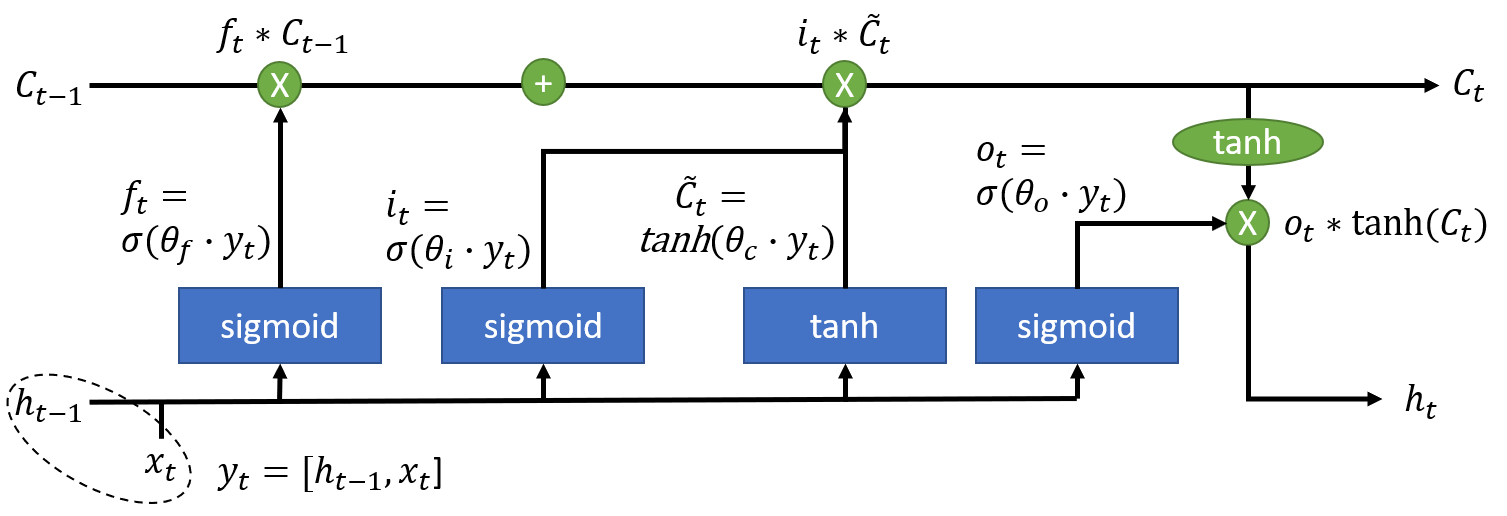}
\vspace{-6mm} 
\caption{LSTM network structure}
\label{fig_LSTM}
\end{figure}

LSTM is a type of recurrent neural network that is capable of learning long-term dependencies, as shown in Fig. \ref{fig_LSTM}. LSTM uses cell state to capture the long-term dependencies of the data, which is determined by three gates, namely the input gate, forget gate, and output gate. LSTM is adopted in our study to learn the temporal correlation of the voltage observations without manually stacking a certain number (kind of feature engineering) to reduce the dimension of the inputs (compared with FNN) and thus accelerating the algorithm. After the LSTM layer, fully connected neural network layers are added, as shown in Fig. \ref{fig_FNN_LSTM}(b).


\subsection{Comparison with existing DRL methods}
{\color{blue} 
\comment{Comparison in terms of scalability, sampling efficiency, convergence, learning rate, sensitivity of hyperparameters, on-line v.s. off-line, parameter space vs action space exploration}

Here we summarize the distinguished features of the proposed PARS algorithm compared with other DRL algorithms as follows.
\begin{enumerate}
\item PARS is highly scalable. Over 90\% of the computation can be easily parallelized. For this reason, the learning speed almost scales linearly with the number of CPU cores. This feature is particularly important for applications that require huge amounts of environment simulation computation, such as reinforcement learning for power grid emergency, where the dynamics simulation of large-size power grid takes more than 90\% of the computations time. In our next section, we used over 500 CPUs for training, achieving a speed up of 136 times compared to its series implementation. Although researchers have devoted significant efforts to parallelize other DRL algorithms, such as the state-of-the-art PPO algorithm that we compared with in the paper in the next section, it is much difficult to have a fully parallel implementation because there are complex dependencies between different components of these algorithms \cite{liang2018rllib}, and thus only a small portion of the algorithms can run in parallel. Consequently, as shown in Section \ref{subsec:IEEE300}, with the same amount of computational resources, our PARS algorithm completes the training five times faster than the parallel PPO. 

\item PARS is a more robust learning algorithm. One of the main shortcomings of many DRL algorithms is their sensitivity to the choice of hyperparameters and random seeds \cite{irpan2018deep, henderson2017deep}. Considerably manual tuning is required to make these methods work. In contrast, PARS has only five hyperparameters, as discussed in Section \ref{subsec:ars_intro}. Even better, PARS is less sensitive to hyperparameters and random seeds. As shown in Section \ref{subsec:IEEE300}, we kept the hyperparameters fixed, and PARS learned consistently high-performing policies for different random seeds. Contrarily,  the parallel PPO algorithm had large variance in the training performance, and failed to learn satisfactory control policies more frequently.

\item PARS does not need back-propagation. Most DLR algorithms use a forward pass to rollout the policy, and a back-propagation pass to calculate the policy gradient. Back-propagation is a computational intensive operation for deep neural networks, and usually demands for small step-sizes in learning. Since PARS is a sampled-based optimization solver, it only needs the forward pass, and estimates the policy gradient via evaluating multiple forward passes. As a result, PARS is often 2-3 times faster in gradient estimation in practice \cite{salimans2017evolution}, and supports comparatively larger step-sizes in learning.

\item PARS performs structured and more effective exploration. Exploration is critical for DRL to find optimal policies. Most DRL algorithms adopt action-space exploration, for example, by adding Gaussian noise to the output action of the neural network policy. This exploration can cause a random jitter on spot, search only locally, and is not an effective exploration strategy for long-horizon decision making. In contrast, PARS uses parameter-space exploration. Instead of perturbing the actions randomly, it perturbs the policies. Parameter-space exploration was shown to lead to more consistent exploration, a richer set of behaviors, and more successful training \cite{plappert2017parameter}.

\end{enumerate}

One weakness of our PARS method is that it requires at least a modest parallel computing environment (e.g., with 10+ CPU cores) to perform well. However, this is not a bottleneck as such a computing setup is quite common now, even in laptops.

}

\section{Test Cases and Results}\label{sec:results}

\comment{
{\color{red}
TODO for R1:
\begin{enumerate}
\item clarify real-time performance requirement of emergency control
\item clarify parallel contribution: novel architecture
\item Comparison with other existing methods
\item Hierarchical structure increases sampling efficiency (?? test case to verify
\item add action mask
\item Compare with MPC, Ray APPO or DD APO ( show performance after certain time steps; normalize the time baseline
\item expand future work
\end{enumerate}}
}

Our ARS training framework is deployed on a local high-performance computing cluster with a Linux operating system which comprises 520 nodes. Each node features dual-socket Intel Haswell E5-2670V3 CPU (12 cores per socket running at 2.3 GHz) with 64 GB DDR4 memory. We tested the performance of our PARS algorithm with different numbers of computing nodes and cores. Tests were performed with both IEEE 39-bus and IEEE 300-bus systems. {\color{blue}The data of both test systems and trained NN models are publicly available in \cite{RLGC}.}

{\color{blue}
\subsection{Baseline methods}

\begin{enumerate}
\item UVLS: it is one widely used rule-based method in the industry nowadays.
The following 3-stage settings are used in this paper: 1) $V_{th} = 0.70 \: pu, \tau = 0.33 \: s, \Delta P  = 20\% P_{init}$; 2) $V_{th} = 0.80 \:pu, \tau = 0.5 \:s, \Delta P  = 20\% P_{init}$; 3) $V_{th} = 0.90 \: pu, \tau = 1.5 \: s, \Delta P  = 20\% P_{init}$; where $P_{init}$ denotes the pre-fault total load at a load bus.

\item MPC: it is a widely used analytic-based method for dynamic, sequential control problems. A trajectory-sensitivity based MPC \cite{amraee2011adaptive} is chosen for comparison.

\item Parallel PPO: It is a state-of-the-art DRL method and has been used to solve large-scale problems such as OpenAI Five.  We chose the high-performance parallel PPO implementation in RLlib \cite{liang2018rllib} as a baseline that represents existing model-free RL methods. 
\end{enumerate}
}

\subsection{Performance Metrics}
To evaluate the performance of the developed ARS algorithm with its enhancements, the following metrics were defined and investigated.

\begin{enumerate}
  \item \textit{Metrics for training:} We considered (a) computational time  and (b) convergence rate. The total computational time at each training iteration was recorded and accumulated. Less execution time for each training iteration was an indicator of higher computational efficiency. RL training is considered as converged when its learning curve gets flat with small variations (e.g., 2\%). The convergence rate can be represented by an minimum iteration number at which the average reward reaches a stable value. The smaller the iteration number achieving a stable average reward, the faster the training converges.
  
  \item \textit{Metrics for testing:} (a) The average rewards the trained policy obtained on the testing task set that was different from the training task set; (b) total load shedding amount. {\color{blue}Note that the reward defined in Eq. (\ref{eq: UVLS_reward}) measures the optimality of the load shedding controls based on the factor that the control should shed as little load as possible to recover the system voltage.} The comparison of the rewards between ARS and the baseline methods is presented in the following subsections.

\end{enumerate}

\subsection{Test Case 1: IEEE 39-bus Test System} \label{subsec:IEEE39}

We applied our proposed PARS algorithm on the IEEE 39-bus test system (details of the system can be found in \cite{huang2019loadshedding_DRL}) to learn a closed-loop control policy for applying the load shedding at a load center including buses 4, 7, and 18 to avoid the FIDVR and meet the voltage recovery requirements shown in Fig. \ref{fig:1}. We first trained the ARS algorithm with linear, FNN, and LSTM models for representing the control policy. Observations included voltage magnitudes at buses 4, 7, 8, and 18 as well as the remaining fractions of loads served by buses 4, 7 and 18. For the linear and FNN models, the last 10 recent observations were stacked and used as input for ARS, thus the dimension of the input was 70; while for the LSTM models, there was no need for stacking the observations from previous time steps, and thus the dimension of the input was 7. The control action for buses 4, 7, and 18 at each action time step was a continues variable from 0 (no load shedding) to -0.2 (shedding 20\% of the initial total load at the bus). 

\begin{figure}[t]
\centerline{\includegraphics[scale=0.2]{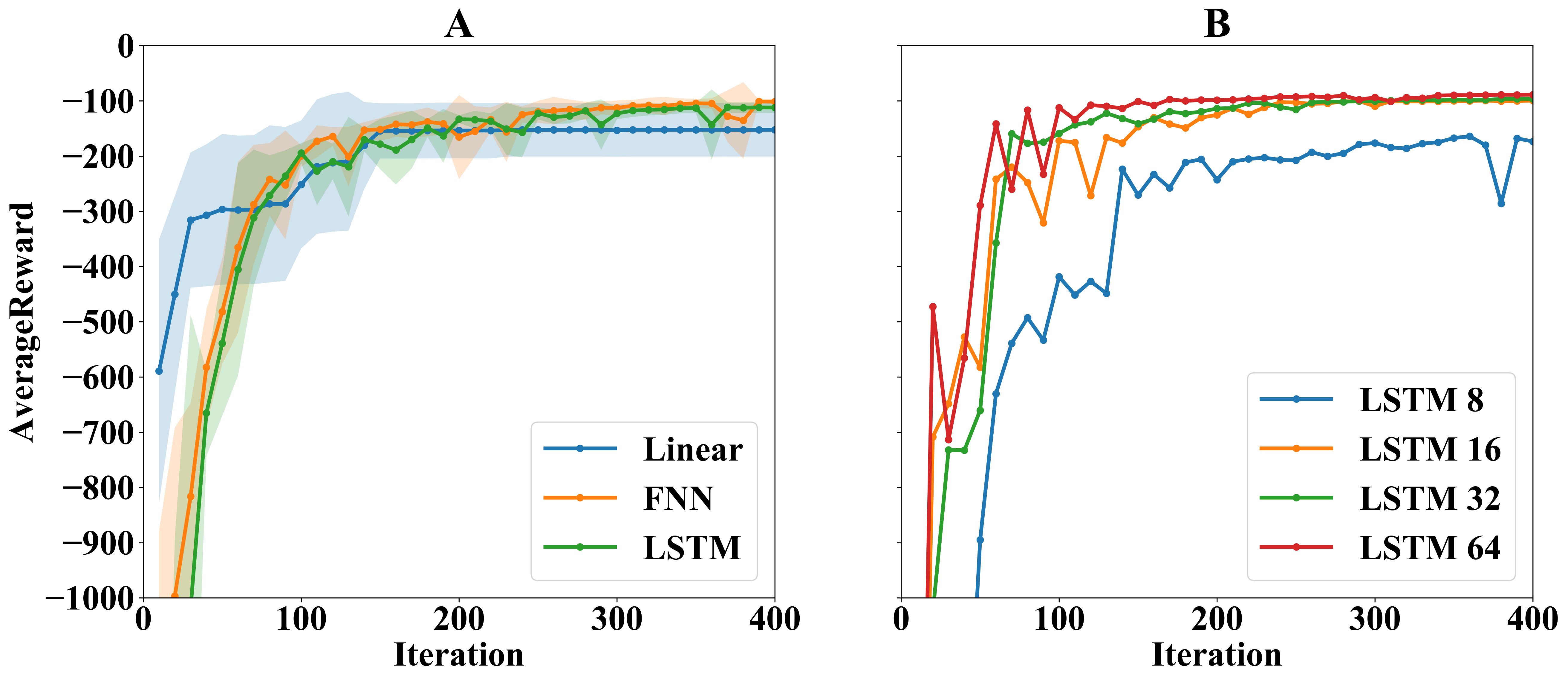}}
\caption{Convergence curves of training using (A) linear, FNN, and LSTM with 16 directions. The training curves were averaged over five random seeds and the shaded region shows the standard deviations; (B) different numbers of directions for LSTM.}
\label{fig_IEEE39_1}
\end{figure}

During the training, the task set $T$ was defined as nine different tasks (fault scenarios). Each task began with a flat start of dynamic simulation. At 1.0 s of the simulation time a short-circuit fault was applied at bus 4, 15, or 21 with a fault duration of 0.0 s (no fault), 0.05 s, or 0.08 s and the fault was self-cleared. The task set $T$ defined with multiple fault locations and durations could guarantee the PARS algorithm interacted with the system with and without FIDVR conditions. Other parameter settings for the PARS algorithm are listed in Table \ref{tab_IEEE39_1}. With the setting of 16 directions for the policy-level perturbation and 9 tasks for task-level domain randomization, the proposed two-level parallelism scheme needed a minimum of 144 cores for fully parallelizing the computation tasks. As a result, eight computational nodes with maximum available cores of 192 were used for training. Fig. \ref{fig_IEEE39_1}(A) shows the reward achieved by PARS using the three different policy architectures, averaged over the five random seeds. The linear model has the lowest performance among the three models. The major performance difference between the FNN and LSTM models is the computational efficiency:  the LSTM model helped reduce the total time by 50\% compared with the FNN model, due to the smaller input dimension (from 70 to 7), while the convergence curves of both models behave in a very similar way as shown in Fig. \ref{fig_IEEE39_1}(A).  

\begin{table}[t]
\caption{Hyperparameters for training IEEE 39-bus system and 300-bus system}
\begin{center}
\begin{tabular}{|l|c|c|c|}
\hline
\textbf{Parameters} & \textbf{39-Bus} & \textbf{300-Bus} \\
\comment{
\hline
Policy Model & Linear, FNN, LSTM \\}
\hline
Policy Network Size (Hidden Layers) & [32,32] & [64,64] \\
\hline
Number of Directions ($N$) & 16 & (32,64,128) \\
\hline
Top Directions ($b$) & 8  & (16,32,64) \\
\comment{
\hline
Number of Fault Cases ($F$)& 9 & 9\\}
\comment{
\hline
No. of Maximum Iterations ($I$) & 500 & 500 & 700\\}
\comment {
\hline
Rollout Length & 150 & 150 \\}
\hline
Step Size ($\alpha$) & 1 & 1\\
\hline
Std. Dev. of Exploration Noise ($\upsilon$) & 2 & 2 \\
\hline
Decay Rate ($\varepsilon$) & 0.99 & 0.996 \\
\hline
\end{tabular}
\label{tab_IEEE39_1}
\end{center}
\end{table}

Based on the above evaluation, we chose the LSTM model to test the parallel scalability of the PARS algorithm, which measured the capacity of effectively using an increasing number of processors. The following two groups of different parallel parameters were investigated:
\begin{itemize}
  \item number of perturbation directions $N$
  \item number of CPUs
\end{itemize}

\begin{figure}[t]
\centerline{\includegraphics[scale=0.2]{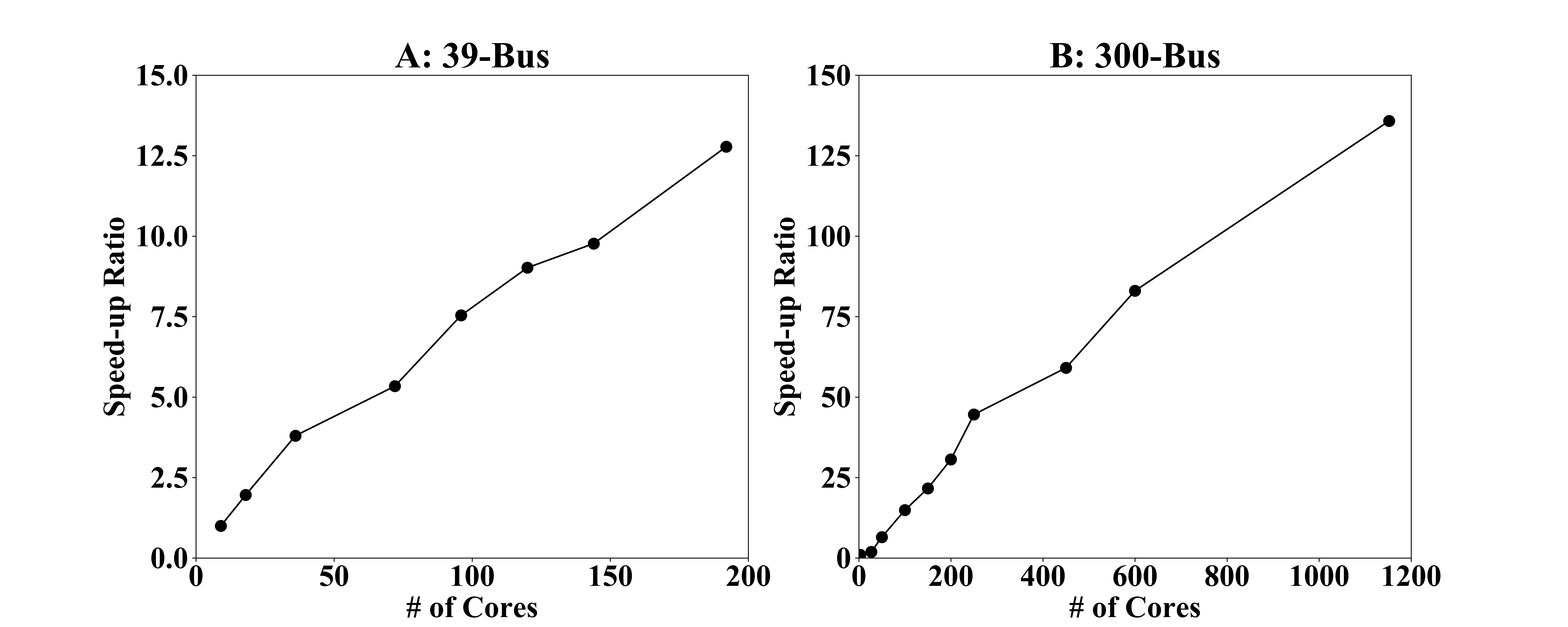}}
\caption{Speed-up ratio using different number of cores. (A) IEEE 39-bus, 300 iterations, LSTM model, 16 directions, 9-core time cost as base case; (B) IEEE 300-bus, 500 iterations, LSTM model, 128 directions, 3-core time cost as base case.}
\label{fig_IEEE39_3}
\end{figure}

The influence of increasing the number of policy perturbation directions $N$ on training performance is significant, as shown in Fig. \ref{fig_IEEE39_1}(B). Using only eight directions might not be sufficient to archive an acceptable performance, while 16 directions are required to reach optimal training results. Using more directions than 16 will improve the convergence rate but requires more computational resources. The parallel scaling performance with different cores for 16 policy perturbation directions is plotted in Fig. \ref{fig_IEEE39_3}(A). It shows that the high-performance computing platform has excellent scalability. The training time is about 5.5 hours for PARS using only 9 cores of one CPU, whereas the training time for DQN is 21 hours in our previous work\cite{huang2019loadshedding_DRL}, demonstrating the high efficiency of PARS. 

\begin{figure}[t]
\centerline{\includegraphics[scale=0.2]{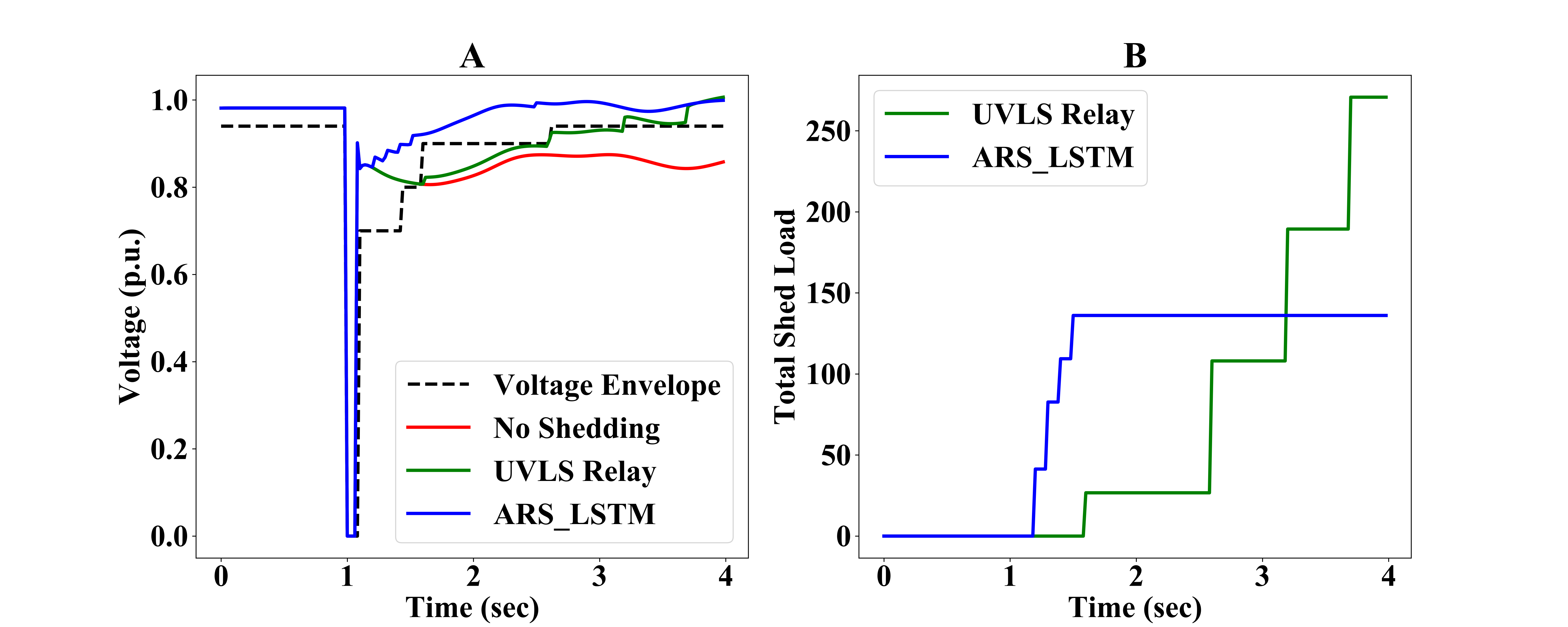}}
\caption{Testing results of trained LSTM 16 model. (A) Voltage of bus 4. Dash line denotes the performance requirement for voltage recovery. (B) Total load shedding amount.}
\label{fig_IEEE39_4}
\end{figure}

We tested the trained LSTM policy on a set of 120 tasks (fault scenarios) with the combination of 30 different fault locations (bus 1 to bus 30) and 4 fault durations (0.02, 0.05, 0.08, and 0.1 s). We also compared the trained PARS-based load shedding control versus the conventional UVLS load shedding scheme. The comparison results show that PARS outperformed the UVLS for all the tasks that required load shedding, as the rewards PARS obtained were always higher than UVLS for those tasks. As a result, either PARS shed less load or UVLS could not recover the system voltage within the required time period to meet the standard defined in Fig.  \ref{fig:1}. Fig. \ref{fig_IEEE39_4} shows the performance comparison between PARS and UVLS for a test task with 0.08 s of fault at bus 3. The total rewards of the PARS and UVLS relay control in this test case were -94.09 and -2367.21, respectively. From Fig. {\ref{fig_IEEE39_4}}(A), it is shown that the voltage with UVLS control (green curve) at bus 4 could not recover within required time to meet the standard (dashed black curve), while the voltage with PARS control (blue curve) could recover to meet the standard. More importantly, Fig. {\ref{fig_IEEE39_4}}(B) shows that the better voltage recovery for PARS is achieved with even less (about 100 MW) total load shedding amount, compared with the UVLS relay control, demonstrating the adaptiveness advantage of PARS over UVLS. 

\subsection{Test Case 2: IEEE 300-bus system} \label{subsec:IEEE300}

\begin{figure}[t]
\centering
\includegraphics[scale=0.25]{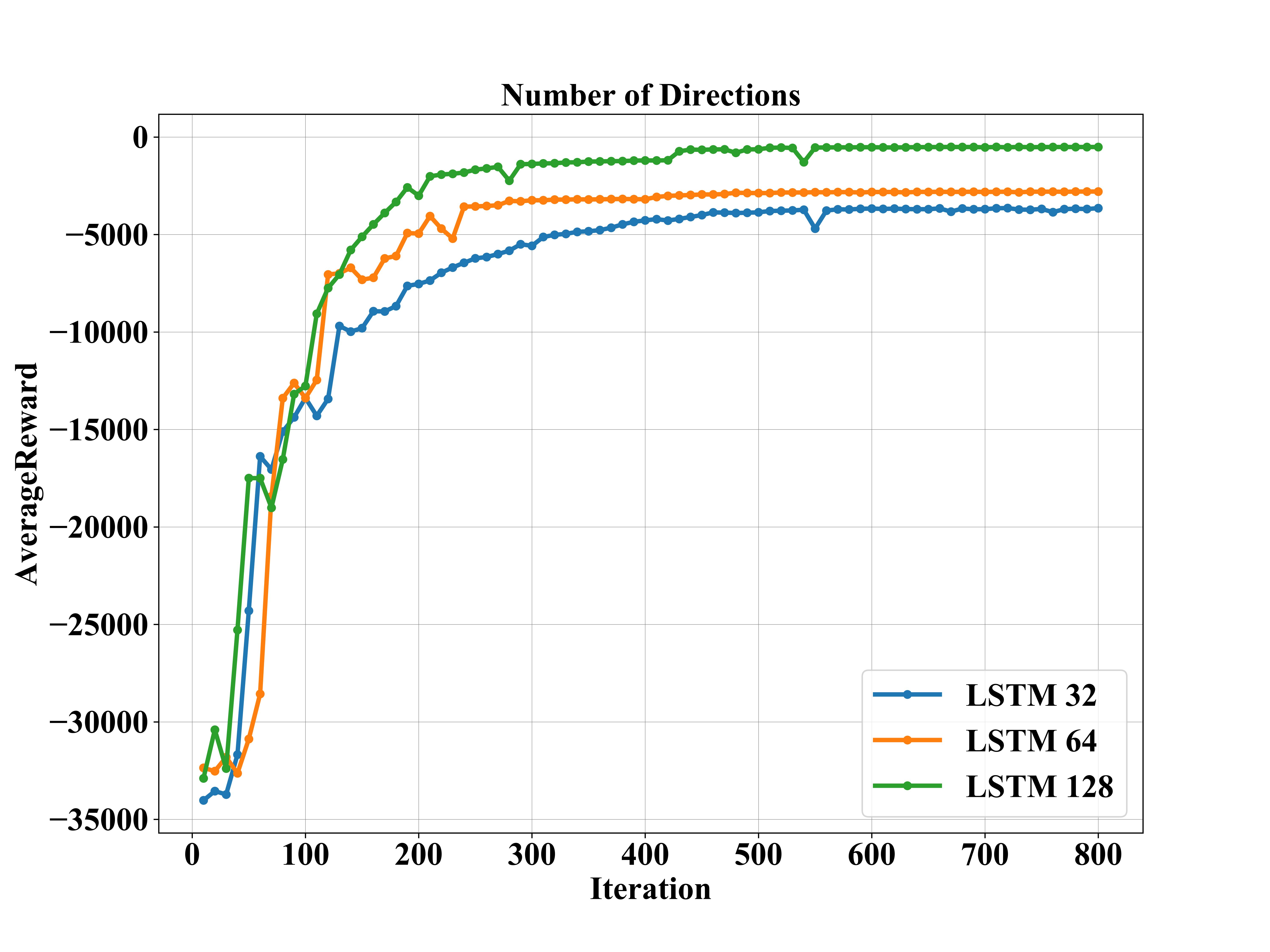}
\caption{Convergence curves of training using different numbers of  directions for the 300-bus system.}
\label{300_converge_diff_directions}
\end{figure}

Based on the 39-bus system training and testing results, LSTM was chosen to model control policy for the IEEE 300-bus system. The possible load shedding control actions were defined for all buses with dynamic motor loads at zone 1 (46 buses in total). The amount of load could be shed for each bus at each action time step is a continuous variable from 0 (no load shedding) to 0.2 (shedding 20\% of the initial total load at the bus). The observations included voltage magnitudes at buses in zone 1 (total 154 buses) as well as the fractions of loads served at the 46 buses where load shedding could be applied; thus the dimension of the input observation was 200. The task set $T$ was defined as 27 different tasks (fault scenarios), which was a combination of 3 fault duration times (0.0, 0.05, and 0.08 s) and 9 candidate fault buses (i.e., 2, 3, 5, 8, 12, 15, 17, 23, 26). At each of the training iterations, 3 fault locations are sampled and combined with 3 potential fault durations to create the rollout tasks. {\color{blue}The hyperparameters of the PARS algorithm for the training are shown in Table \ref{tab_IEEE39_1}}. We trained the LSTM policy with 32, 64, and 128 policy perturbation directions. Fig. \ref{300_converge_diff_directions} shows the average rewards with respect to training iterations under different policy perturbation directions and it was clear that the training converged faster and achieved better final rewards with an increased number of perturbation directions. Fig. \ref{fig_IEEE39_3}(B) shows the parallel scaling performance on the 300-bus system training with different cores for 128 policy perturbation directions. The proposed PARS algorithm gains a speed-up of 136 times when scaling the training from 3 cores to 1152 cores. More importantly, the speed-up curve is almost linear and shows no saturation for up to 1152 cores, suggesting potential for further acceleration when necessary.

\begin{table}[t]
\caption{Hyperparameters of the parallel PPO algorithm different from default values}
\begin{center}
\begin{tabular}{|l|c|}
\hline
\textbf{Parameters} & \textbf{300-Bus} \\
\hline
Number of rollout worker actors & 500 \\
\hline
Learning rate & 5e-4 \\
\hline
Entropy coeff. (Regularizer) & 1e-4 \\
\hline
Number of hidden layers for fully connected net & [64,64] \\
\hline
Clip parameter & 0.3 \\
\hline
Use KL loss & True  \\
\hline
VF clip parameter & 200 \\
\hline
\end{tabular}
\label{tab_PPO}
\end{center}
\end{table}

\begin{figure}[t]
\centering
\includegraphics[scale=0.2]{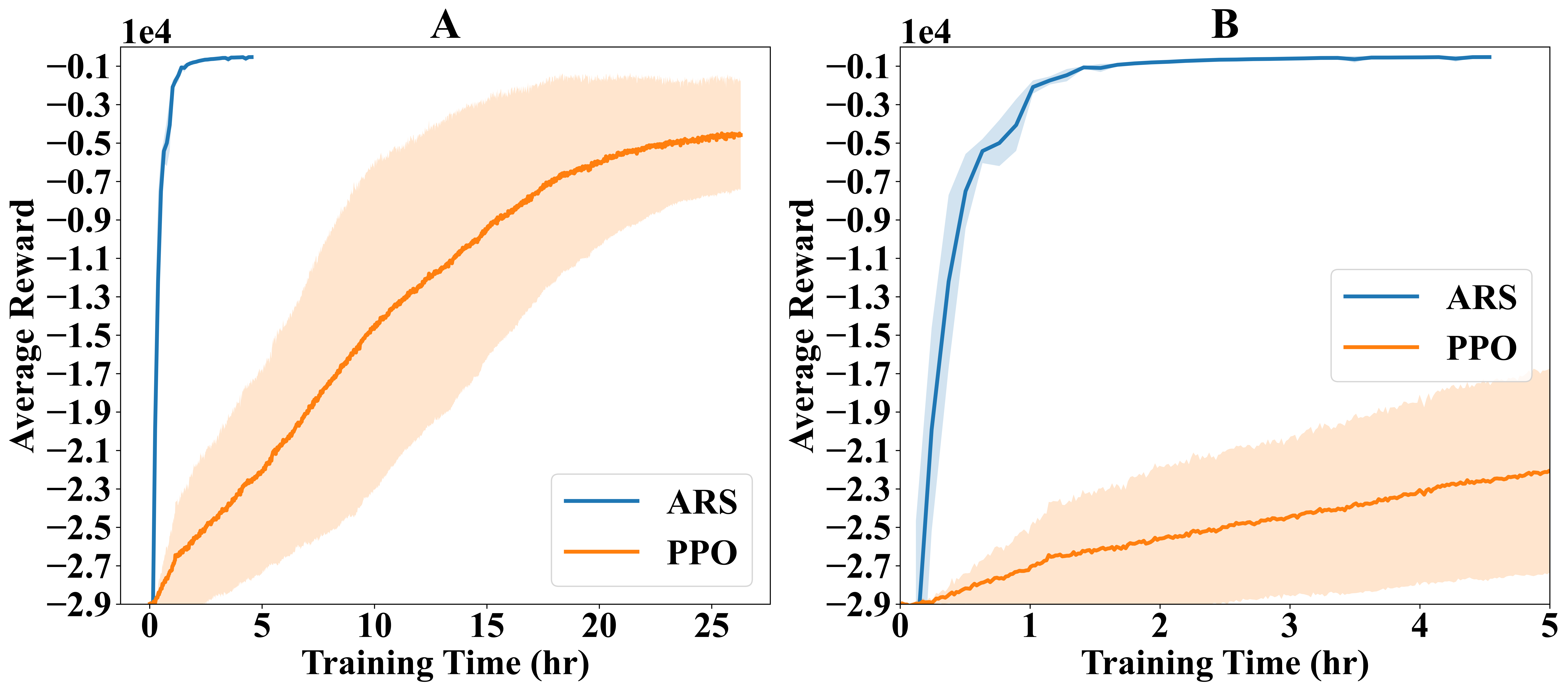}
\caption{Convergence curves of training by time (hr) using PARS and PPO for the 300-bus system for (A) 25 hours (B) 5 hours. The training curves were averaged over five random seeds and  the  shaded  region  shows  the  standard  deviation.}
\label{PPO-1}
\end{figure}

\begin{figure}[t]
\centering
\includegraphics[scale=0.25]{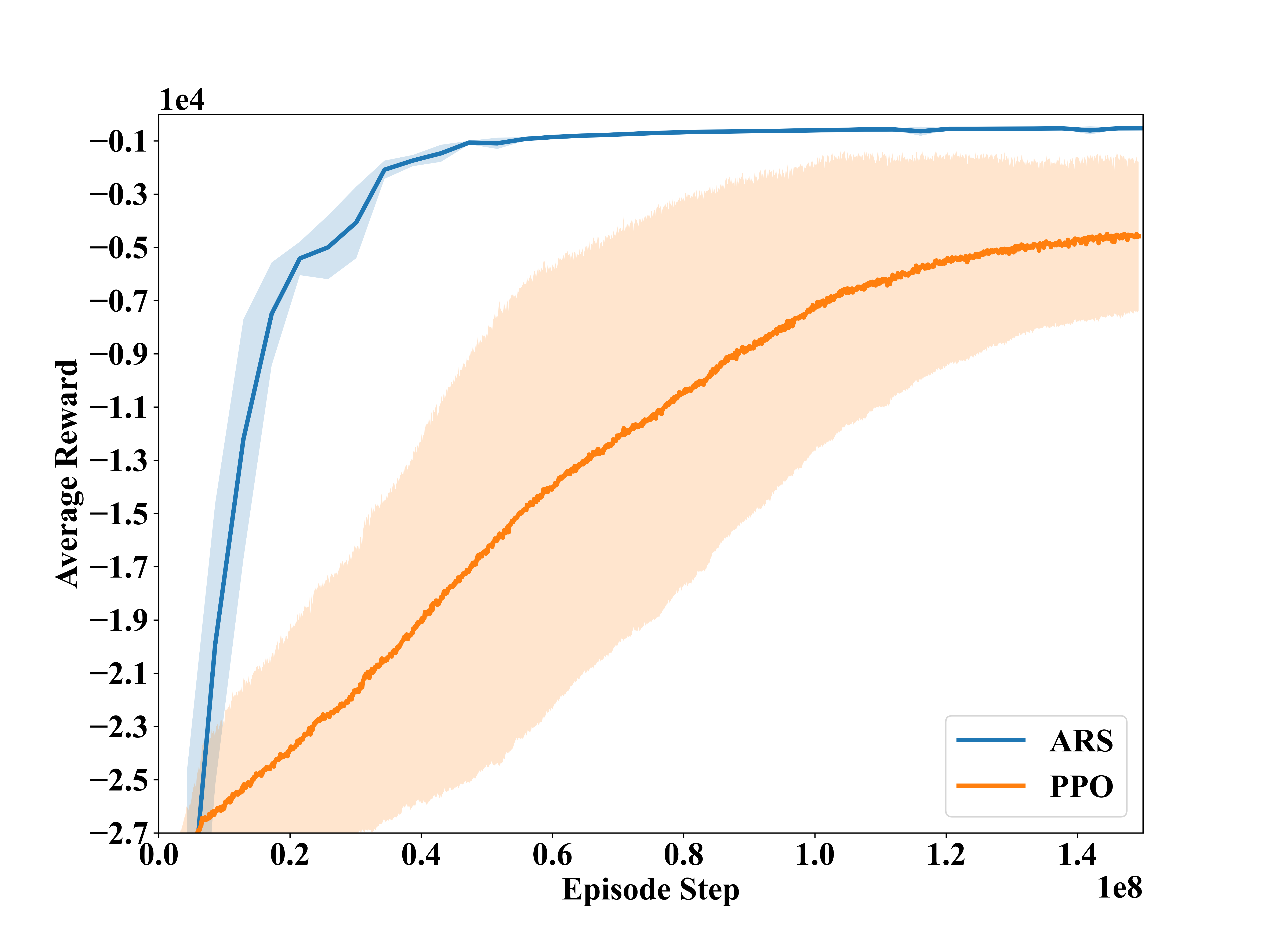}
\caption{Convergence curves of training by episode steps using PARS and PPO for the 300-bus system. The training curves were averaged over five random seeds and  the  shaded  region  shows  the  standard  deviation.}
\label{PPO-2}
\end{figure}

{\color{blue}
We ran the training with PARS and parallel PPO using fixed hyper-parameters but different random seeds for five times. Table {\ref{tab_PPO}} listed hyperparameters of PPO that are different from the default values provided in the RLlib implementation. Both trainings of PARS and PPO use the same computational resources of 500 CPU cores. Fig. \ref{PPO-1} shows the learning curves for both PARS and PPO, with mean and variance, with respect to the wall-clock training time. Fig. \ref{PPO-1} shows that with the same amount of computational resources, our PARS algorithm runs five times faster than the parallel PPO algorithm. Furthermore, our PARS algorithm learns consistently high-performing policies for all the five training cases (the narrow shaded area indicates small variance). In contrast, the performance of PPO has a large variance for the five training cases (the orange shaded region is much wider than that of PARS), and the converged average reward of PPO is lower than that of PARS. Fig. \ref{PPO-2} shows the learning curve for both PARS and PPO, with mean and variance, with respect to the environment simulation steps. Fig. \ref{PPO-2} clearly shows that PARS outperforms the PPO in terms of convergence rate in learning. With the same amount of generated data from simulation environment (the same episode steps), PARS achieved notably higher converged average rewards, as well as better successful rate of training (higher converged average rewards and much smaller rewards deviations). The results demonstrate that for the studied grid voltage emergency control problems, our proposed PARS method achieves much better performance than the state-of-the-art parallel PPO method.
}

After training, we tested the LSTM policy trained with 128 perturbation directions on a set of 170 different tasks (fault scenarios) which correspond to combinations of 34 different fault buses in zone 1 and 5 fault duration times (0.05, 0.06, 0.07, 0.08, and 0.1 s). We also compared the PARS-based load shedding control with the conventional UVLS scheme, the best trained PPO policy, and the solutions obtained by the MPC method.

\begin{figure}[t]
\centerline{\includegraphics[scale=0.2]{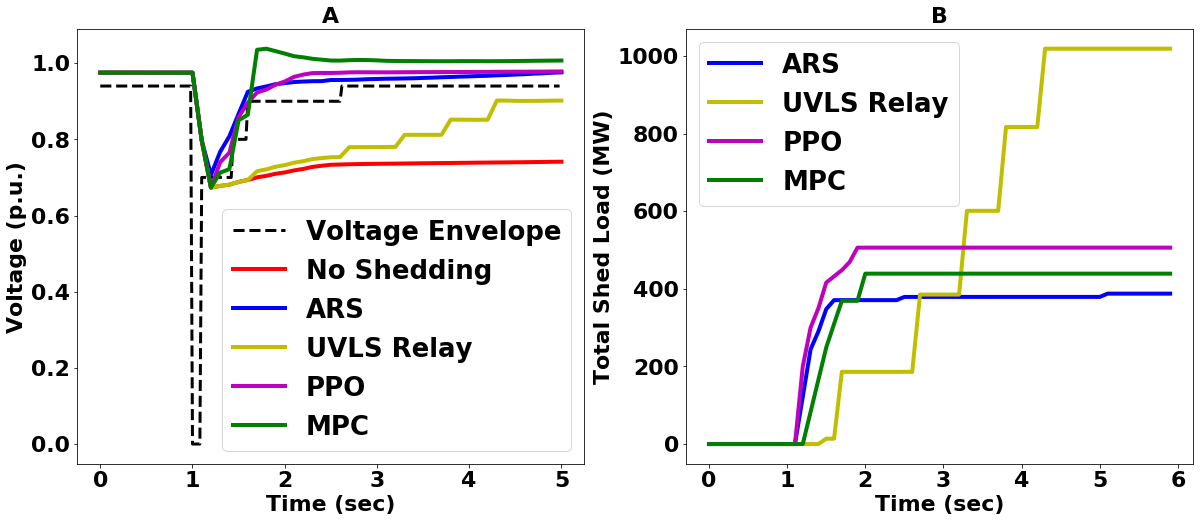}}
\caption{Test results for a FIDVR event triggered by a three-phase fault at bus 23 of the IEEE 300-bus system. (A) Voltage of bus 33. The dash line denotes the performance requirement for voltage recovery. (B) Total load shedding amount.}
\label{300-voltage-load}
\end{figure}

{\color{blue}To show the comparison results, we calculated the reward differences (i.e., the reward of PARS subtracts that of a baseline method) for all the test cases. A positive value means our PARS method performs better for the corresponding test scenario with respect to the optimality defined by Equ. (\ref{eq: UVLS_reward}) and vice versa. Fig. \ref{300_reward_diff} shows the histogram of the rewards differences. As can be seen, PARS outperforms both UVLS and PPO since most of the reward differences are positive when compared with UVLS (98.82\%) and PPO (95.29\%). However, it should be noted that PPO actually has a much better performance when compared with ULVS. Fig. \ref{300_reward_diff} (A) shows that for more than 50\% test tasks, PARS achieves a reward difference more than 10,000 compared with UVLS, which indicates that for these tasks, ULVS fails to recover the voltages back to meet the requirements and thus get a large ``failure'' penalty. In contrast, there are only 3 out of 170 tasks where PPO fails, and Fig. \ref{300_reward_diff} (C) shows that for about 85\% of the test tasks, the reward difference between PARS and PPO are less than 5000. Fig. \ref{300_reward_diff} (B) shows that PARS and MPC achieve comparable results. Table {\ref{tab:table2}} shows the average computation time of the PARS and MPC methods. The computation time for UVLS relays is not included as it is either instantaneous or a predefined delay. It is obvious that PARS requires much shorter solution time than MPC, because performing NN forward pass to infer actions in the PARS approach is much more efficient compared to a solving a complex optimization problem with the MPC method. With 0.72 s action time during a 8-second FIDVR event, the PARS method can meet the real-time operation requirements and allows grid operators to verify the control actions when necessary.

\begin{figure}[t]
\centering
\includegraphics[width=\columnwidth]{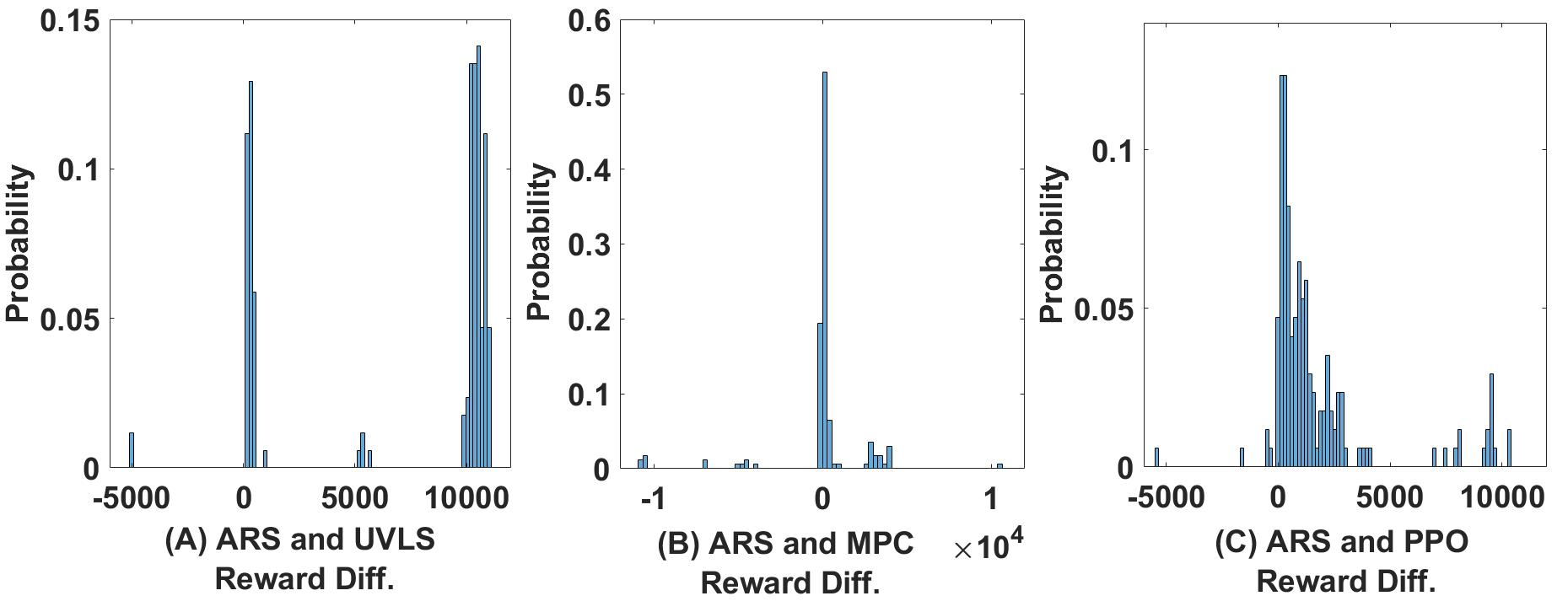}
\caption{Histogram of the reward differences: (A) PARS and UVLS; (B) PARS and MPC; (C)  PARS and parallel PPO.}
\label{300_reward_diff}
\end{figure}

 \begin{table}[t]
 \renewcommand{\arraystretch}{1.2}
 \caption{Comparison of average computation time for PARS and MPC}
     \centering
     \begin{tabular}{|c|c|}
     \hline  
       PARS &  MPC  \\
       \hline  
        0.72 seconds  & 63.31 seconds \\
        \hline
     \end{tabular}
          \label{tab:table2}
 \end{table}

\comment{
\begin{figure}[t]
\centering
\includegraphics[width=\columnwidth ]{figures/reward_his.png}
\caption{Histogram of the differences between the rewards by ARS and UVLS. }
\label{reward_diff}
\end{figure}
}
\comment{
\begin{figure}[t]
\centering
\includegraphics[width=\columnwidth ]{figures/reward_his3.png}
\caption{Histogram of the differences between the rewards by ARS and UVLS. }
\label{reward_diff}
\end{figure}
}

Fig. \ref{300-voltage-load} shows the comparison of PARS, UVLS, PPO, and MPC performance for a test case with a three-phase fault at bus 23 lasting 0.1 second. The total rewards of PARS, MPC, PPO, and UVLS methods in this test case are -823.58, -876.61, -927.34, and -21901.80, respectively. Fig. {\ref{300-voltage-load}}(A) shows that the voltage with UVLS control (green curve) at bus 33 could not recover to meet the standard (dashed black curve), while with PARS (blue curve), PPO (purple curve), and MPC (green curve) methods, voltages could be recovered to meet the performance standard. Further investigation indicated there were five buses that could not recover their voltage with the UVLS control, while the PARS, PPO, and MPC methods could successfully recover voltage at all buses above the required levels. Fig. {\ref{300-voltage-load}}(B) shows that PARS shed the least amount of loads among the four methods. 
}

\section{Conclusions}\label{sec:conclusions}

As described in this paper, a highly scalable DRL algorithm named PARS was developed based on the ARS algorithm and tailored for power system voltage stability control using load shedding. The derivative-free nature and inherent parallelism in the ARS algorithm are fully exploited in PARS.  

PARS is developed on the Ray framework and synergistically integrated with the RLGC platform to achieve high scalability and applicability for power system stability control applications. Furthermore, both FNN and LSTM are considered for policy modeling in PARS to better handle high non-linearities in power systems and enhance its generalization capability to unseen scenarios. A small number of hyperparameters makes PARS easy to tune to achieve good performance. Case studies demonstrated that LSTM performs better than FNN in terms of computational efficiency and that PARS scales almost linearly up to more than 1000 CPU cores with no scaling performance saturation. The high scalability of PARS enables reducing the training time of IEEE 300-bus system from about 40 days to less than 10 hours (i.e., 100X speedup). {\color{blue}The test results on the IEEE 300-bus system showed: 1) compared with MPC, PARS has a notable advantage in the solution time while achieving comparable results; 2) compared with the state-of-the-art parallel PPO method, PARS has strengths in faster convergence while achieving better solutions, as well as robustness to random seeds in training.}

{\color{blue}
There are several important future research directions: 
\begin{enumerate}
    \item The results in this paper show good potential for PARS to be applied in larger power systems and for different control applications  such as frequency control\cite{Yan_Frequency_DRL} and cascading event mitigation\cite{almassalkhi2014MPC_cascade}.
    \item In order to better understand the strength and weakness of different DRL methods and their applicability for various grid control problems, it is important to comprehensively benchmark them through open platforms and standardized test cases.
    \item  As power systems constantly change,  ensuring the  adapability  of  the  machine  learning  model  is  a  requirement for practical acceptance of machine learning applications\cite{Wehenkel2020_ML4RM}. One potential approach is to incorporate meta-learning into DRL to exploit past learning experience to quickly adapt to new operation conditions or learn new tasks\cite{yu2020MSO}.
    \item  Power grid is a mission-critical infrastructure, thus one main concern of application of machine learning (including DRL) for controlling such a critical system is safety and security. Safe RL has been an important research topic and most of previous efforts in this area focus on safe exploration during training\cite{dalal2018safeRL,garcia2015safe_RL}. Whereas in power system, the training is mostly done in simulators and the real concern is safety and robustness of the DRL-based control after deployment. Thus, it is important to investigate and solve safety- and robustness-related issues in DRL-based grid control.
\end{enumerate}
}



\comment{
\appendices
\section{Proof of the First Zonklar Equation}
Appendix one text goes here.

\section{}
Appendix two text goes here.

\section*{Acknowledgment}

The authors would like to thank Dr. Kory Hedman from U.S. DOE ARPA-E for his support and guidance.

}

\ifCLASSOPTIONcaptionsoff
  \newpage
\fi



%

\bibliographystyle{IEEEtran}
\bibliography{ADRL4LS.bib}

%
\comment{
\begin{IEEEbiography}{Michael Shell}
Biography text here.
\end{IEEEbiography}

\begin{IEEEbiographynophoto}{John Doe}
Biography text here.
\end{IEEEbiographynophoto}


\begin{IEEEbiographynophoto}{Jane Doe}
Biography text here.
\end{IEEEbiographynophoto}




}

\end{document}